\documentstyle[aps]{revtex}
\input{epsf.tex}
{\makeatletter%
\gdef\labeleqs#1{{%
\edef\@currentlabel{%
\ifappendixon\appletter\fi
\ifsecnumbers\ifnum\c@secnum>0 
\arabic{secnum}.\fi\fi\arabic{equation}}%
\label{#1}%
}}%
} 
\def\QCD{{\em QCD\,}}
\def\vev{{\em vev\ }}
\def\ee{{\rm e}}
\def\ii{{\rm i}}
\twocolumn
\begin{document}
\draft
\widetext
\title{A disorder parameter for dual  
superconductivity in gauge theories.
}
\author{A. Di Giacomo and G. Paffuti}
\thanks{
Partially supported by MURST and by EC
Con\-tr\-act 
CHEX-CT92-0051.
}
\address{Dipartimento di Fisica dell'Universit\`a and 
I.N.F.N., I-56126 Pisa, Italy
{\centerline{e-mail:digiacomo@pi.infn.it}}
{\centerline{e-mail:paffuti@ipifidpt.difi.unipi.it}}
}
\maketitle
\widetext
\begin{abstract}
A detailed discussion is given of the disorder parameter for dual superconductivity
of lattice gauge theories, introduced in a previous paper, and of its relation to
other approaches. New lattice data are reported. Among other results, we find that
the dual superconductivity of compact $U(1)$ is type II.
\end{abstract}
\setcounter{page}{1}
%
\narrowtext
\section{Introduction}
Dual superconductivity of the vacuum is an important phenomenon in gauge
theories.
\begin{itemize}
\item[-] It produces confinement of electric charges via dual Meissner effect in
the abelian case.
\item[-] It is most likely the mechanism of colour confinement in \QCD\cite{1,2}
\item[-] It plays a key role in the structure of supersymmetric gauge
theories\cite{3}.
\end{itemize}
The simplest case is the compact $U(1)$ gauge theory.

With Wilson action this theory shows a phase transition at $\beta_c\simeq
1.01$, probably weak first order, from a phase at low $\beta$ where electric
charge is confined, to a phase of free photons\cite{4}.

Confinement is detected by measuring the string tension from the vacuum
expectation value (\vev) of Wilson loops. The penetration depth of the
electric field is finite for $\beta < \beta_c$, indicating dual Meissner effect,
and goes large at the deconfining transition $\beta_c$\cite{4}.

Monopoles are detected by their Dirac strings as units of $2\pi$ magnetic flux
through the plaquettes. Their number density is not a disorder parameter for
dual superconductivity, in the same way as the number of electric charges is
not for ordinary superconductivity.
However, empirically, the number density of monopoles is larger in the confined
phase, and drops to zero above $\beta_c$\cite{4}.

A legitimate disorder parameter should vanish for symmetry reasons in the
deconfined phase, and be different 
from zero in the confining phase. Since dual
superconductivity is nothing but the spontaneous breaking of the $U(1)$
symmetry related to the magnetic charge conserva\\

\vphantom{a}
\vskip7\baselineskip\noindent
tion, the \vev of any operator
carrying non zero magnetic charge can be a disorder parameter.A non zero \vev
of
such an operator would indeed indicate that vacuum has
not a definite
magnetic charge, i.e. that monopoles condense in it in the same way as Cooper
pairs do in the ground state of ordinary superconductors.
The concept of disorder parameter is known since long time in the comunity of
field theory and statistical mechanics\cite{5,6}. In the pionering numerical
simulations in  lattice gauge theories, however, emphasis was given to the
density of monopoles as indicators of dual superconductivity\cite{4,7}.

A rigorous proof was given in ref.\cite{8} that monopoles condense at low
$\beta$'s in lattice $U(1)$ theory with Villain action\cite{9}. The proof makes
use of the specific form of the action and so did numerical attempts to extract
a disorder parameter\cite{10}. However, probably because of the mathematical
language of the forms, which is not so familiar to physicists, nobody tried for
long time to export the construction to the generic form of the action, or to
non abelian gauge theories.
Indeed, after abelian projection\cite{11}, monopole condensation in non abelian
gauge theories like \QCD, always reduces to an effective $U(1)$ with
Dirac monopoles\cite{11,12}.
Of course the $U(1)$ effective action is unknown, and therefore a construction
of the disorder parameter is needed, which can work with any variant action.

Such a construction was given in ref\cite{13} and immediately afterwards was
used to demonstrate dual superconductivity of non abelian theories\cite{14}.

This result prompted the exportation of the construction of ref.\cite{8,10}
from Villain to generic action.

In this paper we want to discuss in detail and improve the construction of
ref\cite{13} (sect.2), compare it to that of ref.\cite{8}, (sect.3), 
showing that they are equivalent,
and
present a number of numerical results for lattice $U(1)$ with Wilson's action,
(sect.4). We will then compare our\cite{13} way of detecting superconductivity by the
quantity $\rho = \frac{d}{d \beta}\ln\langle\mu\rangle$, $\langle\mu\rangle$
being the disorder parameter, to direct determination of $\langle\mu\rangle$ or
of its effective potential (sect.4).

Besides confirming dual superconductivity of $U(1)$ gauge theory
in the confined phase we show that it is II kind.

The present discussion is also a useful basis to the treatment of the analogous
problem in non abelian theories, which will be presented elsewhere.

\section{The disorder parameter.}
The construction of ref\cite{13} of the creation operator of a monopole or
antimonopole, is inspired by ref\cite{5,6} and is based on the following simple
idea.

In the Schr\"odinger representation where the field $\vec A(x)$ is diagonal, a
monopole of charge $2\pi\frac{q}{e}$ sitting in $\vec y$ is created by adding the
corresponding vector potential $\frac{1}{e}\vec b(\vec x - \vec y\,)$ to $\vec
A(x)$.

This is nothing but a translation of $\vec A(x)$, which is generated by the
conjugate momentum $\vec\pi(x) = \vec E(x)$, the electric field operator.
In the same way as
\begin{equation}
\ee^{\ii pa} | x\rangle = |x +a \rangle \label{eq:2.1}\end{equation}
we have
\begin{equation}
|\vec A(\vec x,t) + \frac{1}{e}\vec b(\vec x-\vec y)\rangle =
\mu  |\vec A(\vec x,t) \rangle \label{eq:2.2}\end{equation}
with
\begin{equation}
\mu(\vec y,t) = \exp\left[\ii \frac{1}{e}\int {\rm d}^3 x\,\vec E(\vec x,t)
\vec b(\vec x-\vec y)\right] \label{eq:2.3}\end{equation}
The magnetic charge operator being
\begin{equation}
Q = \int {\rm d}^3 x \vec\nabla\left(\vec\nabla\wedge \vec A(\vec x,t)\right)
\label{eq:2.4}\end{equation}
the commutator $[Q,\mu]$ can be evaluated by use of the canonical commutation
relations
\begin{equation}
\left[ E_i(\vec x,t), A_j(\vec z,t)\right] = -\ii \delta_{ij} \delta^3(\vec x -
\vec z) \label{eq:2.5}\end{equation}
giving
\begin{eqnarray}
\left[ Q(t), \mu(\vec y,t)\right] &=&
\frac{1}{e} \int {\rm d}^3 x 
\vec\nabla\left(\vec\nabla\wedge \vec b(\vec x -\vec y)\right)\cdot\mu(\vec
y,t) \nonumber\\
&=& \frac{q}{2 e} \mu(\vec y,t) \int {\rm d}^3 x \vec\nabla\left(\frac{\vec
r}{r^3}\right) = \nonumber\\
&=&2\pi \frac{q}{e}\mu(\vec y,t)\label{eq:2.6}
\end{eqnarray}
In deriving Eq.(\ref{eq:2.6}) the Dirac string has been removed. 

A choice for $\vec b(\vec x-\vec y)$ can be
\begin{equation}
\vec b(\vec x-\vec y) = \frac{q}{2}
\frac{\displaystyle \vec r\wedge \vec n_3}{\displaystyle r(r - \vec r\cdot
\vec n_3)} \label{eq:2.7}\end{equation}
Alternative choices differ by a gauge transformation, $\vec b\to \vec b + \vec
\nabla\Phi$ which leaves the operator invariant if 
the Gauss law
$\vec\nabla\cdot\vec E = 0$
is satisfied.

On the lattice the building block of the theory is the link $U_\mu(n)$, which
is an element of the gauge group. For $U(1)$ $U_\mu(n) = \ee^{\ii \theta_\mu(n)}$
and the plaquette, $\Pi_{\mu\nu}$, which is the parallel transport along the
elementary square in the plane $\mu \nu$ at the site $n$, is
\begin{equation}
\Pi_{\mu\nu}(n) = \exp(\ii \theta_{\mu\nu}(n))\label{eq:2.8}\end{equation}
with
\begin{equation}
\theta_{\mu\nu} = \Delta_\mu \theta_\nu - \Delta_\nu \theta_\mu
\mathop\simeq_{a\to 0} a^2 e F_{\mu\nu}\label{eq:2.9}\end{equation}
The lattice version of the electric field is then
\begin{equation}
a^2 E_i \simeq \frac{1}{e} {\rm Im}\,\Pi^{0i} + {\cal O}(a^4)
\label{eq:2.10}\end{equation}
and a definition of the operator $\mu$ on the lattice\cite{13} can be
\begin{eqnarray}
\mu(\vec y,n_0) &=&
\exp\left[-\beta \sum_n b^i(\vec n - \vec y) {\rm Im} 
\Pi^{0i}(\vec n,n_0)\right] = \nonumber\\
&=& \exp\left[-\beta  \sum_{\vec n} b^i(\vec n -\vec y) \sin(\theta_{0i}
(\vec n,n_0))\right]
\label{eq:2.11}\end{eqnarray}
$\beta = 1/e^2$.
Here $b^i(\vec n)$ is the discretized version of the monopole field 
Eq.(\ref{eq:2.7}).
The factor $\beta$ in front of the exponent comes from the factor $\frac{1}{e}$
in the monopole charge
times the normalization factor in Eq.(\ref{eq:2.10}). Usual Wick
rotation to Euclidean region has been performed.

The form (\ref{eq:2.11}) was successfully used in ref.\cite{13}.

A better definition of $\mu$ can be given, which coincides with
Eq.(\ref{eq:2.11}) in practice, but automatically respects the compactness of
the theory, in that it shifts the exponent of the links, and not the links
themselves. In formulae:
\begin{eqnarray}
\mu(\vec y,m_0) &=&
\exp
\Big\{ 
\beta\sum_{\vec n}
\big( 
\cos[ \theta^{0i}(\vec n,m_0) + b^i(\vec y - \vec n)] - 
\nonumber\\
&-&
\cos[ \theta^{0i}(\vec n,m_0) ] \big)\Big\}
\label{eq:2.12}\end{eqnarray}
For small $b^i$ the definition (\ref{eq:2.12}) coincides with (\ref{eq:2.11}).

More generally if $\sum_{\mu\nu n} S(\theta_{\mu\nu}(n))$ is the action, 
$\mu$ will
be defined as
\begin{eqnarray}
\mu(\vec y,m_0) &=& \exp\Big\{
\beta\sum_{\vec n}\big[ S(\theta^{0i}(\vec n,m_0) + b_i(\vec n-\vec y) ) -
\nonumber\\
&-&
 S(\theta^{0i}(\vec n,m_0) )\big]\Big\}
\label{eq:2.12b}\end{eqnarray}
and will tend to the expression (\ref{eq:2.11}) as the lattice spacing $a$ go to zero,
when the action tends to the continuum action.

The prescription of excluding Dirac string on a lattice being either to locate
the monopole at $\vec y$ between two neighbouring sites, or to eliminate in the sum
the arrow of sites where $\vec b$ is singular, it is easy to verify that the
definitions (\ref{eq:2.12}) and (\ref{eq:2.11}) give the same results from the
practical point of view.

If the action is the Wilson's action\cite{15}
\begin{equation}
S = \sum_{n,(\mu\nu)} \beta\left( \cos(\theta_{\mu\nu}) -
1\right)\label{eq:2.13}\end{equation}
then the vacuum expectation value of $\mu$ is given by
\begin{equation}
\langle \mu \rangle = \frac{1}{Z}\int \left[\prod_{\mu,n}\,{\rm
d}\theta_\mu(n)\right]\,\exp(S)\,\mu\label{eq:2.14}\end{equation}
or, making use of (\ref{eq:2.12})
\begin{equation}
\langle \mu(\vec y,m_0) \rangle = \frac{1}{Z}\int \left[\prod_{\mu,n}
\,{\rm d}\theta_\mu(n)\right]\,\exp(S + S')\label{eq:2.15}\end{equation}
where $S'$ is the exponent of Eq.(\ref{eq:2.12}).

Adding $S'$ simply amounts to modify the $(0,i)$ plaquet\-tes on the time slice
$n_0$, by addition of $b_i$ to $\theta_{0i}$
\begin{eqnarray}
S+S' &=&
\sum_{n}\sum_{(i,j)=1}^3\beta \left(\cos(\theta_{ij}(n) - 1\right) +
\label{eq:2.16}\\
&+&
\sum_{n,n_0\neq m_0}
\beta \left(\cos(\theta_{0i}(n) - 1\right) +\nonumber\\
&+&\sum_{\vec n}
\left(\cos(\theta_{0i}(\vec n,m_0) + b^i(\vec m-\vec n))-1\right)
\nonumber
%
\end{eqnarray}
If a number of monopoles and antimonopoles are created at time $n_0$, $b_i$
should be the sum of the corresponding vector potentials. The generic
correlation function $\langle \mu(x_1)\ldots \mu(x_n)\rangle$ is defined as
$\langle\mu\rangle$ in Eq.(\ref{eq:2.15}), with the change from $S$ to $S+S'$ extended
to all the time slices where monopoles or antimonopoles are created.

So for example the correlation function where a monopole is created in $\vec y =
0$ at $t=0$ and destroyed at $t=n_0$ is given by
\[ \langle \mu(\vec y,0)\,\bar\mu(\vec y, m_0) \rangle =
\frac{1}{Z}\int \exp(S+S'_{\mu\bar\mu})\]
$S+S'$ differs from $S$ by the replacement
\begin{eqnarray}
\theta_{0i}(\vec n,0) &\to& \theta_{0i}(\vec n,0) + b_i(\vec n - \vec y)
\qquad{\rm at}\,t=0 \nonumber\\
\theta_{0i}(\vec n,m_0) &\to& \theta_{0i}(\vec n,m_0) - b_i(\vec n - \vec y)
\qquad{\rm at}\,t=m_0 \label{eq:2.18}\end{eqnarray}
Monopole condensation can be detected from the asymptotic value of $\langle
\mu(\vec y,0)\,\bar\mu(\vec y, n_0) \rangle$. Indeed as $n_0$ grows large, by
cluster property
\begin{equation}
\langle\mu(\vec y,0)\,\bar\mu(\vec y, m_0) \rangle\simeq
C\,\exp(-m_0 M) + \langle \mu\rangle^2\label{eq:2.19}\end{equation}
Notice that $\langle\mu\rangle = \langle\bar\mu\rangle$ by $C$ invariance, and the
position $\vec y$ is irrelevant by translation invariance. $M$ is the mass of
the lowest state with monopole charge $q$ in units of inverse lattice spacing.

To visualize that $\mu$ really creates a monopole at $t=0$ consider again the
change it produces according to Eq.(\ref{eq:2.18}). Since
\begin{equation}
\theta_{0i}(\vec n,0) = \theta_i(\vec n,1) - \theta_0(\vec n,0) -\theta_0(\vec
n + \hat i,0) + \theta_0(\vec n,0)\label{eq:2.20}\end{equation}
the change (\ref{eq:2.18}) of $\theta_{0i}$ can be considered as a shift
\begin{equation}
\theta_i(\vec n,1)\to \theta_i(\vec n,1) - b_i(\vec n - \vec
y)\label{eq:2.22}\end{equation} 
A change of variables 
\begin{equation}
\theta'_i = \theta_i(\vec
n,1) - b_i(\vec n - \vec y)
\label{eq:2.20a}\end{equation}
in the Feynman integral (\ref{eq:2.15}), 
which leaves the measure invariant,
brings back the plaquette
$\theta_{0i}$  to its unperturbed form. However the change of variables
(\ref{eq:2.20a}) changes the $(i,j)$ plaquette at $n_0 = 1$ as follows
\begin{equation} \theta_{ij}(\vec n,1)\to
\theta_{ij}(\vec n,1) + \Delta_i b_j(\vec n - \vec y) - \Delta_j b_i(\vec n
-\vec y)
\label{eq:2.20b}\end{equation}
This means that at $n_0=1$ the magnetic field of a monopole located at $\vec n = \vec y$ is
added to the original configuration. The change of
variables (\ref{eq:2.22}) also affects the plaquette $\theta_{0i}(\vec n,2)$, and
amounts to the shift
\begin{equation}
\theta_{0i}(\vec n,2) \to \theta_{0i}(\vec n,2) - b_i(\vec n -\vec y)
\label{eq:2.20c}\end{equation}
Again a change of variables $\theta_i(\vec n,2)\to \theta_i(\vec n,2) -
b_i(\vec n-\vec y)$ restores $\theta_{0i}(\vec n,2)$ to the initial form at the
price of adding a monopole at time $t=2$, and of producing a shift in the form 
(\ref{eq:2.20c}) on $\theta_{0i}(\vec n,3)$. This procedure can be iterated.
 At $t=m_0$ this procedure ends,
because $b_i$ cancels with the shift of opposite sign corresponding to the
creation of the antimonopole.

Thus the correlator $\langle\mu(\vec y,0)\,\bar\mu(\vec y, m_0) \rangle$ simply
consists in having a monopole propagating in time, from $0$ to $n_0$.

The construction above simply generalizes to more complicated forms of the
action, where Wilson loops other than plaquettes enter.

\section{Comparison with other approaches}
In this section we want to discuss the relation of our approach to that of
ref.\cite{8}.

In the language of ref.\cite{8} $\theta_\mu(n)$ is a 1 form associated to the
links and ${\rm d}\,\theta$ is the two form associated to the plaquettes, or the
field strength tensor.

In this language
the partition function is
\begin{equation}
Z = \int{\cal D}[\theta]\,\Phi_\beta({\rm d}\theta)\label{eq:3.1}\end{equation}
For Wilson's action
\begin{equation}
\Phi_\beta = \exp(\beta\sum_{plaq}\left(\cos({\rm d}\theta) - 1\right)
\label{eq:3.2}\end{equation}
For Villain's action
\begin{equation}
\Phi_\beta = \sum_{n}\exp\left\{-\frac{\beta}{2}\sum_{plaq} \Vert {\rm d}\theta
+ 2\pi n\Vert^2\right\}\label{eq:3.3}\end{equation}
To define a disorder operator $\langle\mu\rangle$  the action is modified by
adding a two form $X$ to ${\rm d}\theta$.
We define
\begin{equation}
Z(X) = \int {\cal D}[\theta]\,\Phi_\beta({\rm
d}\theta + X)\label{eq:3.4}\end{equation}
and
\begin{equation}
\langle \mu \rangle = \frac{\displaystyle Z(X)}{\displaystyle Z(0)} 
\label{eq:3.5}\end{equation}
Any change of $X$ of the form $X\to X+{\rm d}\Lambda$ leaves $\langle\mu\rangle$
invariant, in that ${\rm d}\Lambda$ corresponds to a shift of $\theta$ to $\theta +
\Lambda$ which is reabsorbed by a change of the (periodic) integration
variables.

Since a generic $X$ can be written as (Hodge decomposition):
\begin{equation}
X = {\rm d}\alpha + \delta\frac{1}{\Delta}{\rm d} X \label{eq:3.6}\end{equation}
the above invariance implies that $\langle\mu\rangle$ only
depends on ${\rm d} X$.

${\rm d} X$ is a 3-form $[{\rm d}X]_{\mu\nu\alpha}$ and its dual $*{\rm d}X$ is a 1
form, which is a magnetic current, since $X$ is a field strength.
Explicitely
\begin{equation}
{\rm d}X_{\mu\nu\alpha} = -\left(
\partial_\alpha X_{\mu\nu} + \partial_\mu X_{\nu\alpha} + \partial_\nu X_{\alpha\mu}
\right)\label{eq:3.7}\end{equation}
and
\begin{equation}
J^M_\rho = \frac{1}{6}\varepsilon_{\rho\mu\nu\alpha}\,{\rm d} X_{\mu\nu\alpha}
\label{eq:3.8}\end{equation}
The magnetic current (\ref{eq:3.8}) is identically conserved. In the language of
forms
\begin{equation}
\delta J^M = 0 \label{eq:3.9}\end{equation}
The magnetic charge density which describes the creation of a monopole of charge $2\pi
q$ in the site $\vec y$ at time $y^0$, and its destruction at time $y'^0$ is
\begin{equation}
J_0^M(\vec x,x^0) = 2\pi q\delta^3(\vec x -\vec y)\left(\theta(x^0-y^0) - \theta(x^0
-y'^0)\right)\label{eq:3.10}\end{equation}
Since the current is conserved
\begin{eqnarray}
\vec\nabla {\vec J}^M &=& -\Delta_0 J_0^M = \nonumber\\
&=&
- 2\pi q\delta^3(\vec x -\vec y)\left(\delta(x^0-y^0) - \delta(x^0
-y'^0)\right)\label{eq:3.11}\end{eqnarray}
A solution of Eq.(\ref{eq:3.11}) is
\begin{equation}
{\vec J}^M(\vec x,x^0) = 2\pi q\frac{1}{4\pi}\frac{\displaystyle \vec x -\vec y}{
\displaystyle |\vec x-\vec y\,|^3}
\left[\delta(x^0-y^0) - \delta(x^0 -y'^0)\right]
\label{eq:3.12}\end{equation}
The corresponding $X$ is then
\begin{equation}
\overline X = \delta\frac{1}{\Delta} J^M \label{eq:3.13}\end{equation}
The correlation function of a monopole antimonopole will then be:
\begin{equation}
\langle \mu(\vec y,y_0)\,\overline\mu(\vec y,y'^0)\rangle =
\frac{\displaystyle Z(\overline X)}{Z(0)}\label{eq:3.14}\end{equation}
This is the construction of ref.\cite{8}.

Notice that $Z(X)$ is periodic in $X$ (with period $2\pi$) since the action is
compact. In fact $Z$ only depend on ${\rm d}X$, and is periodic also in ${\rm d}X$
with the same period. This can be rigorously proved by going to Fourier transform:
\begin{equation}
Z( {\rm d} X + 2\pi n) = Z({\rm d} X) \label{eq:3.15}\end{equation}
Consider now a one form $\Omega$ on the dual lattice, with support on a line. If
$\delta \Omega = 0$ the support must be a closed line. If $\Omega$ is integer valued
in units of $2\pi$ the change
\[ {\rm d} X = * J^M\to {\rm d} X = * J^M + \Omega\]
leaves $Z$ invariant.

In the notation of ref.\cite{8} $\vec J^M$ is denoted by $2\pi q B$ and $J_0^M$ by
$-2\pi q\omega$ and
\begin{equation}
{\rm d} \overline X = 2\pi q(B- \omega)\label{eq:3.16}\end{equation}
Any $X$ with the same ${\rm d}X$, will give the same correlation function
(\ref{eq:3.14}). The construction presented in sect.~2 
corresponds to the choice
\begin{eqnarray}
{\overline X}'_{0i} &=& b_i(\vec x)
\left[\delta(x^0-y^0) - \delta(x^0 -y'^0)\right]\nonumber\\
{\overline X}'_{ij} &=& 0\label{eq:3.17}\end{eqnarray}
or, in the dual language
\begin{eqnarray}
(*{\overline X}')_{0i} &=& 0\nonumber\\
(*{\overline X}')_{ij} &=& \varepsilon_{ijk} b_k(\vec x)
\left[\delta(x^0-y^0) - \delta(x^0 -y'^0)\right]
\label{eq:3.18}\end{eqnarray}
and
\begin{equation}
*{\rm d}{\overline X}'_\mu = \delta (*{\overline X}')_\mu = -
\sum_\rho \Delta_\rho (*X)_{\rho\mu}\label{eq:3.19}\end{equation}
Explicitely
\[ \delta (*{\overline X}')_0 = 0\qquad
\delta (*{\overline X}')_i = - \sum_k \Delta_k (*X)_{ki}\]
and by Eq.'s (\ref{eq:3.17}) and (\ref{eq:2.7})
\begin{eqnarray}
\delta (*{\overline X}')_i &=& 2\pi q \frac{1}{4\pi}
\frac{x_j - y_j}{|\vec x-\vec y|^3}\left(\delta(x_0-y_0) -
\delta(x_0- y'_0)\right) \label{eq:3.20}\\
&-&
2\pi q \delta(x_1-y_1)\delta(x_2-y_2)\theta(x_3-y_3)\cdot\nonumber\\
&\cdot&\left(\delta(x_0-y_0) -
\delta(x_0- y'_0)\right)
\nonumber\end{eqnarray}
Our $*{\rm d}{\overline X}'$ differs from $*{\rm d}{\overline X}$ (\ref{eq:3.16})
of
ref.\cite{8} by a 1 form integer valued in units of $2\pi$, with support on a closed
line. Therefore our correlator coincides with that of ref.\cite{8}, not only for
Villain action, but for generic form of the action.

This section is a cultivated way of presenting the argument already given at the end
of last section.

\section{Numerical results for the disorder parameter.}
As discussed in sect. 2, we measure the correlation function
\begin{equation}
{\cal D}(x^0) = 
\langle \mu(\vec x,x^0), \bar\mu(\vec x,0)\rangle \simeq
A {\rm e}^{-M x^0}  + \langle \mu\rangle^2
\label{eq:4.1}\end{equation}
The aim is to extract $\langle \mu\rangle^2$, which will signal dual
superconductivity, and $M$ which is the lowest mass in the sector of
magnetically charged excitations.

A direct determination of ${\cal D}$ can be done, as we will discuss below, but
is rather noisy from numerical point of view. The reason for this is that
${\cal D}$
\begin{equation}
{\cal D} = \frac{1}{Z} \int{\cal D}\theta\,\exp(S + S')
\label{eq:4.2}\end{equation}
is the average of $\exp(S')$, $S'$ being the modification of the action on the
time slices $t=0$ and $t=x^0$, and $S'$ fluctuates roughly like the square root
of the spatial volume.

A way to go around this difficulty is to measure, instead of ${\cal D}$ the
quantity\cite{12}
\begin{equation}
\rho(\vec x,x^0,\vec x,0) = \frac{d}{d \beta}\ln{\cal D}
\label{eq:4.3}\end{equation}

At large distance $(x^0\to\infty)$
\begin{equation}
\rho_{\infty} \simeq 2\frac{d}{d \beta}\ln{\langle\mu\rangle}
\label{eq:4.4}\end{equation}
and since $\rho(\beta = 0) = 1$
$\langle\mu\rangle$ can be reconstructed as
\begin{equation}
\langle\mu\rangle = \exp\left(\frac{1}{2}\int\rho(\beta')d\beta'\right)
\label{eq:4.5}\end{equation}
From Eq.(\ref{eq:4.2})
\begin{equation}
\rho_{\infty} = \langle S\rangle_S - \langle S + S'\rangle_{S+S'}
\label{eq:4.6}\end{equation}
The definition of $\rho$ is analogous to the 
definition of the internal energy in terms of 
the partition function
in statistical mechanics. $\rho$ is now a well defined
quantity and easy to measure, and, as we shall see, can give all the information
needed to detect dual superconductivity.

We have made simulations on a $6^3\times12$, $8^3\times16$ and $10^3\times20$
lattices putting the time axis along the long edge of the lattice. A typical
behaviour of $\rho$ versus $x^0$ is shown in Fig.1, for a $8^3\times16$ lattice,
showing that an asymptotic value is reached by $\rho$ as a function of $x^0$. The mass
$M$ of the exponential in Eq.(\ref{eq:4.1}) can be estimated and is typically
$\sim(2-3)/a$
\par\noindent
\begin{minipage}{0.9\textwidth}
\epsfxsize = 0.5\textwidth
\epsfbox{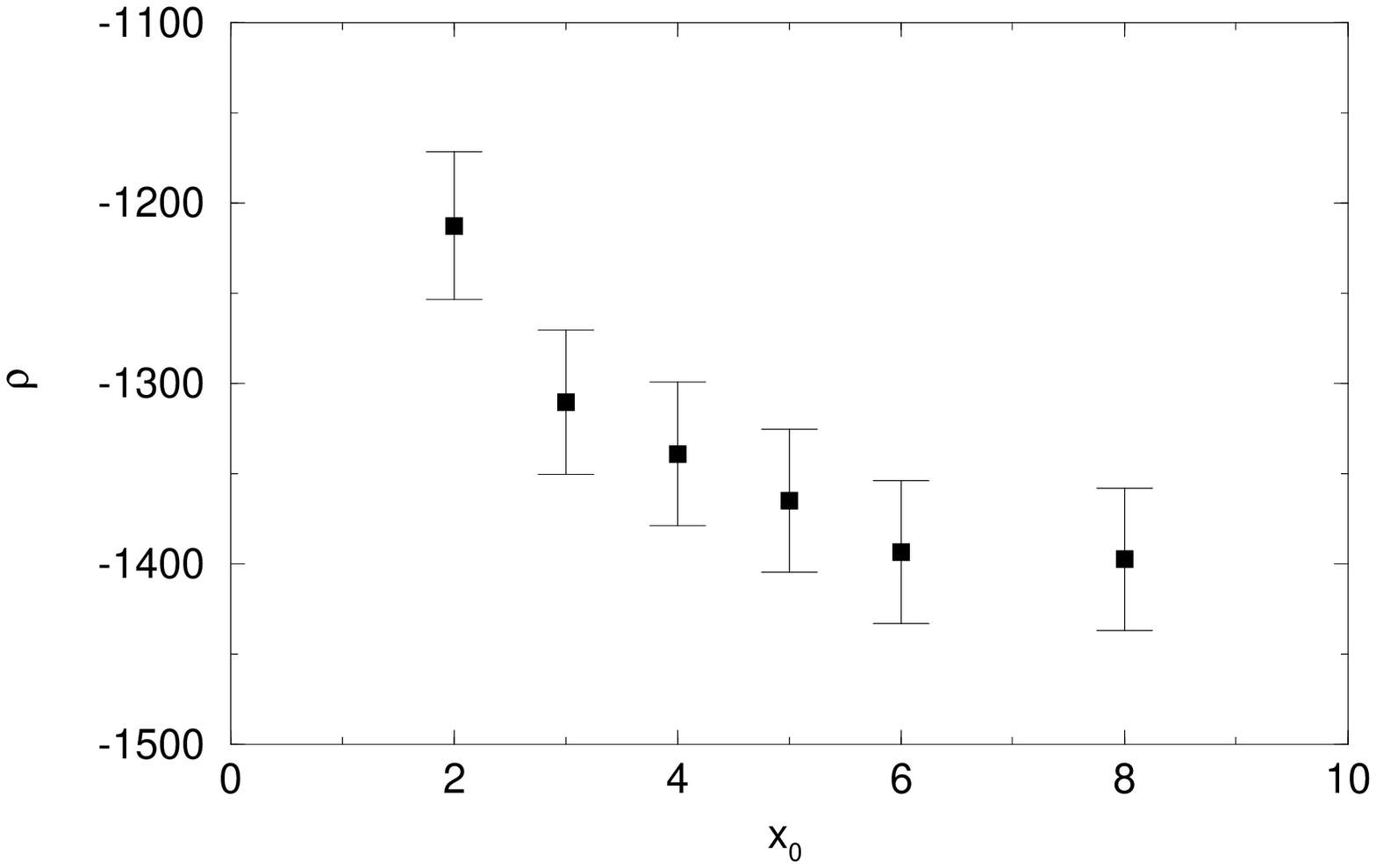}
Fig.1 Monopole antimonopole correlation in time.\\
(Lattice $8^3\times16$)\\
\end{minipage}
\vskip0.1in
We will come back again to this point in the following.

The quantity $\rho_{\infty}$ as a function of $\beta$ is plotted in Fig.2.
\par\noindent
\begin{minipage}{0.9\textwidth}
\epsfxsize = 0.5\textwidth
\epsfbox{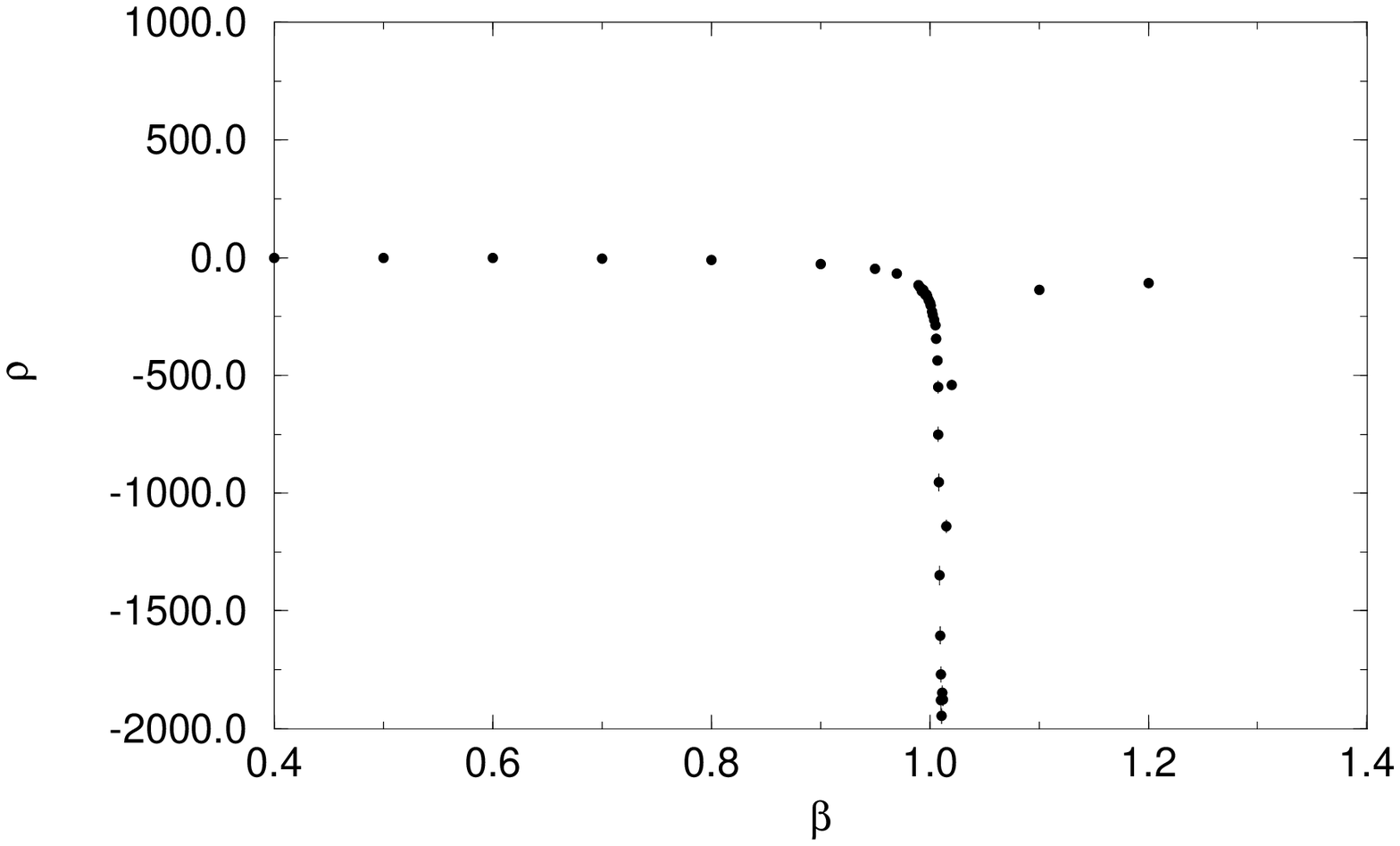}
Fig.2 $\rho_\infty$ as a function of $\beta$. 
The negative peak signals\\
the phase transition.
(Lattice $8^3\times16$)
\end{minipage}
\vskip0.1in
For all of our lattices sizes $\rho_\infty$ is negative and sharply decreases
approaching
$\beta_c$. This corresponds, by Eq.(\ref{eq:4.5}) to a behaviour of 
$\langle\mu\rangle$
which slowly decreases from the value $\langle\mu\rangle=1$ at $\beta = 0$, and
has a sharp drop at
$\beta_c$.

To better analyse this behaviour we compare it for the three lattice sizes
under study. For $\beta < \beta_c$ below the negative peak, $\rho$ increases
with $L$, showing that as $L\to\infty$, $\langle\mu\rangle$ reaches a finite,
nonzero value. 
Magnetic $U(1)$ is therefore spontaneously broken, and for
$\beta < \beta_c$ the system is a dual superconductor (fig.3).

For $\beta \simeq \beta_c$ we know that the typical correlation length of the
system goes large. There is evidence that the transition is weak first
order\cite{16}, with some controversy\cite{16b}.

The correlation length $\xi$ goes large as $\beta$ approaches $\beta_c$ 
in a range of $\beta$'s
and
eventually stops growing before reaching it.
\par\noindent
\begin{minipage}{0.9\textwidth}
\epsfxsize = 0.5\textwidth
\epsfbox{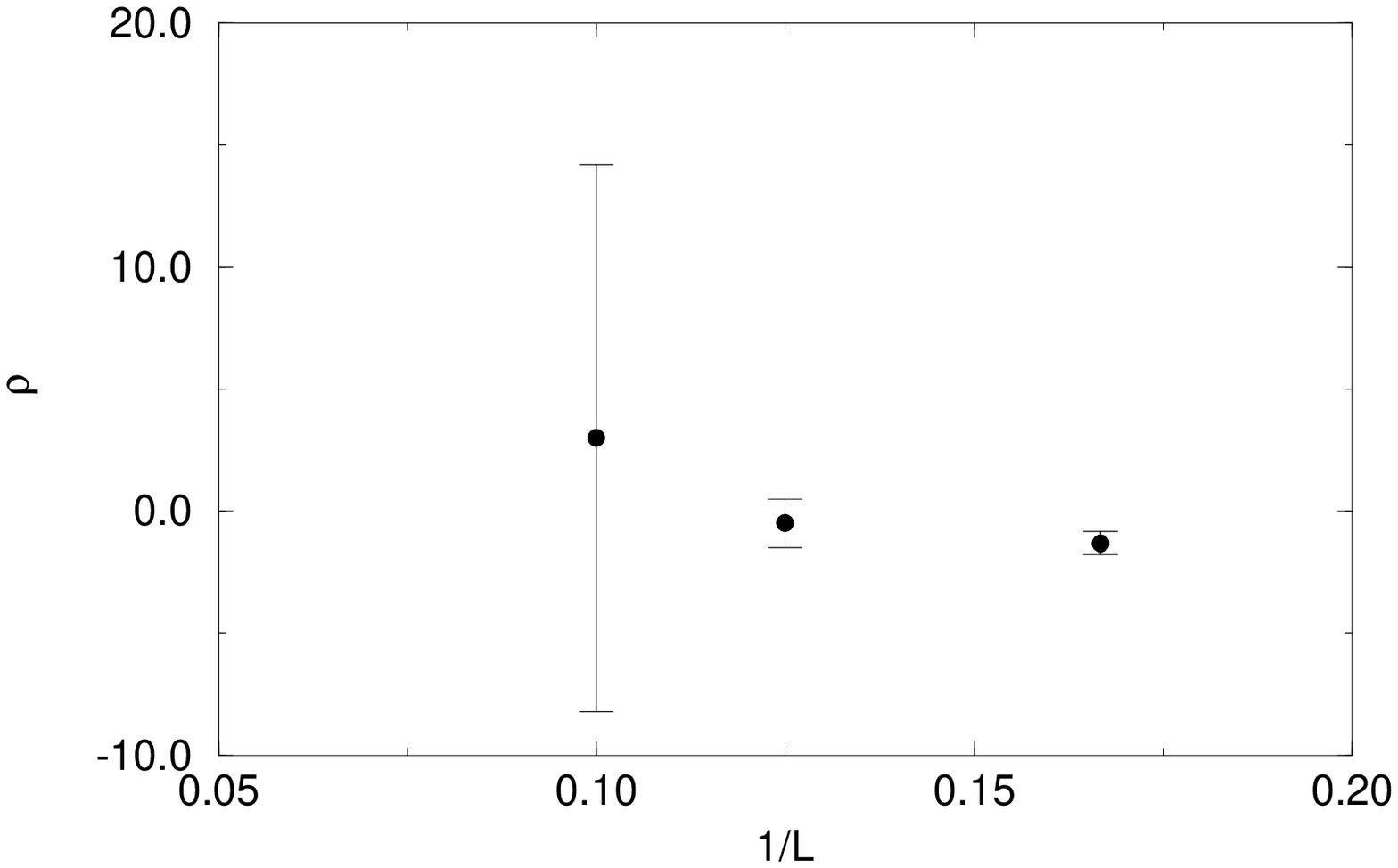}
Fig.3 $\rho_\infty$ versus $1/L$ for $\beta 1.009$. 
\end{minipage}
\vskip0.1in
\noindent
This means that, in the neighbourhood of $\beta_c$
\begin{equation}
\mu  = \mu\left(\frac{\xi}{L},\frac{a}{\xi}\right) \simeq
\mu\left(\frac{\xi}{L}\right)\label{eq:4.7}\end{equation}
If the transition were second order a critical index $\nu$ would exist such that
\begin{equation}
\xi \mathop\simeq_{\beta\to\beta_c^-}\left(\beta_c-\beta\right)^{-\nu}
\label{eq:4.8}\end{equation}
In our case some effective index $\nu$ could anyhow exist, describing
a behaviour of $\xi$ of the form (\ref{eq:4.8}) in the above mentioned
range of $\beta$'s. Then
$\xi/L$ can be traded with $L^{1/\nu}(\beta_c-\beta)$ and a finite size scaling
behaviour results
\begin{equation}
\mu = \mu[ L^{1/\nu}(\beta_c-\beta)]
\label{eq:4.9}\end{equation}
implying for $\rho=\frac{d}{d \beta}\ln\langle\mu\rangle$ a scaling behaviour
\begin{equation}
\frac{\displaystyle \rho}{
\displaystyle L^{1/\nu}} = f\left( L^{1/\nu}(\beta_c-\beta)\right)
\label{eq:4.10}\end{equation}
Eq.(\ref{eq:4.10}) allows a determination of $\nu$ and $\beta_c$, togheter with a
determination of the exponent $\delta$ by which $\langle\mu\rangle$ tends to
zero at
$\beta_c$ in the infinite volume limit.

The quality of the scaling is shown in Fig.4. points corresponding to different
lattice sizes follow the same universal curve only for the appropriate values
of $\beta_c$ and $\nu$, Eq.(\ref{eq:4.11}). If $\beta_c$ or $\nu$ are changed
by one standard deviation from the values of Eq.(\ref{eq:4.11}) points
from different lattices
start
splitting apart from each other.
\par\noindent
\begin{minipage}{0.9\textwidth}
\epsfxsize = 0.5\textwidth
\epsfbox{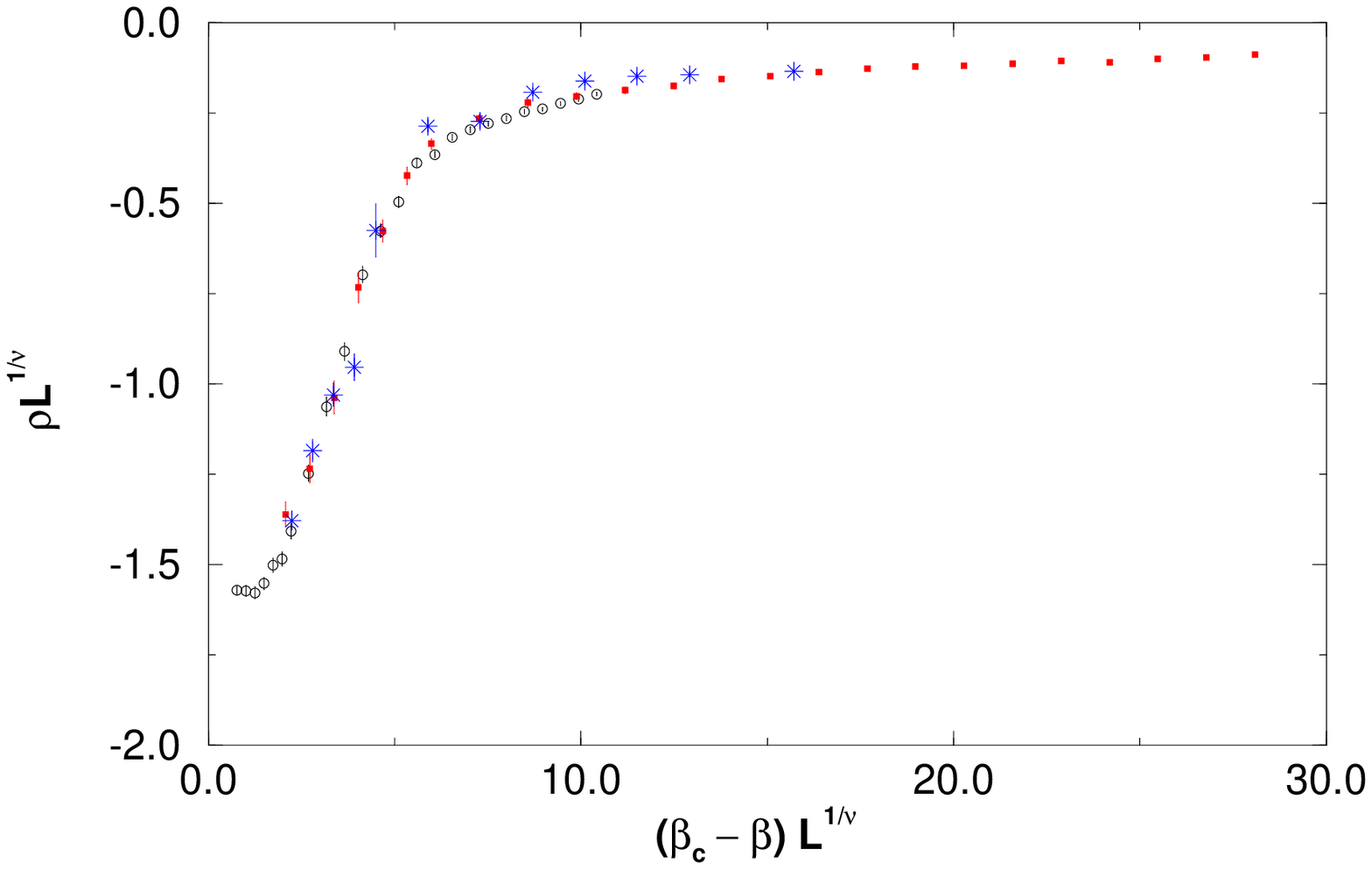}
Fig.4 Finite size scaling. $\rho L^{1/\nu}$ is plotted\\
vs.
$(\beta_c-\beta)L^{1/\nu}$.
\end{minipage}
\vskip0.1in
\noindent
A best square fit gives
\begin{eqnarray}
\beta_c &=& 1.01160(5)\label{eq:4.11}\\
\nu &=&0.29(2) \nonumber
\end{eqnarray}
The value (\ref{eq:4.11}) of $\beta_c$ is consistent with determination
based on completely different methods\cite{16}.
If $\mu\to (\beta_c-\beta)^\delta$
\begin{equation}
\frac{\displaystyle \rho}{
\displaystyle L^{1/\nu}} \simeq
-\frac{\delta }{L^{1/\nu} (\beta_c-\beta)}
\label{eq:4.10b}\end{equation}
An estimate for $\delta$ from the behaviour in fig.4 is:
\begin{equation}
\delta = 1.1 \pm 0.2
\label{eq:4.10cb}\end{equation}
In the region $\beta\to\infty$ $\rho$ can be computed in the weak coupling
approximation\cite{13}. The result is
\begin{equation}
\rho = -5.05\cdot L + 4.771
\label{eq:4.12}\end{equation}
giving $\rho\to-\infty$ or $\langle\mu\rangle = 0$ in the infinite volume limit,
in agreement with general arguments\cite{5}: only as $V\to\infty$ the disorder
parameter vanishes in the disordered phase, if boundary conditions are not free.

The mass of the monopole in Eq.(\ref{eq:4.1}) should scale properly in the
limit $\beta\to\beta_c$ but we have large errors and this behaviour is not clearly
visible (fig.5).

In order to determine if the superconductor is first kind or second kind we
have also measured the penetration depth $1/m_A$ of the electric field on the
lines of ref.\cite{4}\footnote{In ref.\cite{4} the field was called magnetic.}. A
constant electric field parallel to the space boundary of the lattice is put on
a face of the space lattice and its value is determined inside the bulk as a
function of the distance from the boundary. An exponential behaviour is found,
with a penetration depth which properly scales by approaching the critical point,
consistently with the effective critical index.

The corresponding mass is shown in fig.5 together with the mass extracted from
the correlation length, Eq.(\ref{eq:4.1}). It appears clearly that $M \geq
2 m_A$, indicating that the superconductor is second kind.
This same problem has been approached by looking at the Abrikosov flux tubes
generated by propagating charges. The idea is to compare the dependence of the
electric field inside the tube on the transverse distance $x_\perp$ from the
center of the tube, with what is expected from London equations.
Their result is that the system seems to be  is at the border between first and second
kind\cite{17}.
The method is ingenuos. However derivatives are approximated by finite
differences, the penetration depth being a few lattice spacings (2-3), and this can
produce systematic errors. Our method would give a more precise determination if we
were able to determine better the mass $M$ of Eq.(\ref{eq:4.1}).
The question deserves further study.
\par\noindent
\begin{minipage}{0.9\textwidth}
\epsfxsize = 0.5\textwidth
\epsfbox{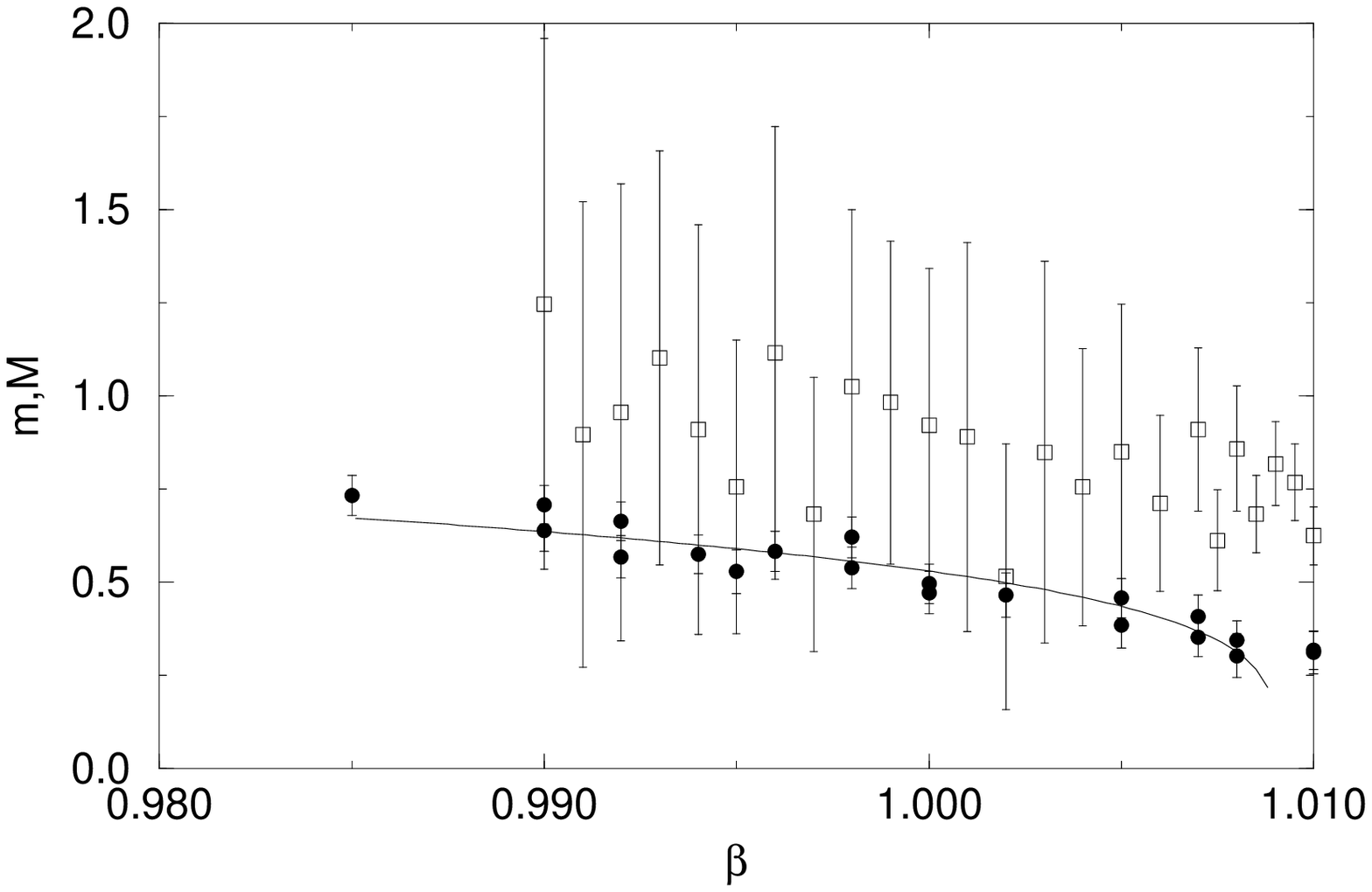}
Fig.5 Mass of the monopole $M$ (squares), and mass of the \\
dual photon 
$m$ (circles) vs. $\beta$.
\end{minipage}
\vskip0.1in
\noindent

Finally we want to comment on the possibility of determining numerically
$\langle\mu\rangle$ directly and not through the measurement of $\rho$. As we
have seen this is not strictly necessary, since $\rho$ gives complete
information about the phase transition. However the problem has some interest
by itself.

The definition of $\langle\mu\rangle$ is $\langle \ee^{\beta S'}\rangle$, the average
being performed with the weight
${\cal D}\theta \ee^{\beta S}/Z$. $S'$ is itself a random variable in this
ensemble which has some average value $\langle S'\rangle$ with a width $\sigma =
\sqrt{\langle S'^2\rangle - \langle S'\rangle^2}$.

A general theorem of probability theory states that if a random variable is
distributed with a probability law $p(x)$, with $\int p(x){\rm d}x = 1$, then its
average $x_n = \frac{1}{n}\sum_k x_k$ is distributed as a gaussian for large $n$ if
and only if\cite{18}
\begin{equation}
\lim_{X\to\infty}\frac{\displaystyle
X^2\int_{|x| > X} p(x){\rm d} x}
{\displaystyle
\int_{|x| < X} x^2 p(x){\rm d} x} = 0
\label{eq:5.1}\end{equation}
If  Eq.(\ref{eq:5.1}) holds, then
\begin{equation}
\langle x_n\rangle\mathop\rightarrow_{n\to\infty} \langle x\rangle = \int\,x\,
p(x) {\rm d} x 
\label{eq:5.1b}\end{equation}
and the width of the distribution is in this limit
\[ \sigma_n = \frac{\sigma}{\sqrt{n}}\]
with
\[\sigma^2 = \langle x^2\rangle - \langle x\rangle^2\]
If we denote by $y$ the variable $\beta S' - \langle \beta S'\rangle$ and by
$\pi(y)$ its probability distribution, then the variable $\mu$
\[ \mu = \exp(\beta S') = \bar\mu exp(y)\qquad 
\left(\bar\mu = exp(\langle \beta S'\rangle)\right)
\]
will be distributed as
\begin{equation}
p(\mu) = \pi\left(\ln\left(\frac{\mu}{\bar\mu}\right)\right)\,{\rm d}
\ln\left(\frac{\mu}{\bar\mu}\right)
\label{eq:5.2}\end{equation}
If $\pi$ decreases as $\exp(-y^2/2\sigma_y^2)$ as $y\to\infty$ then the probability
distribution (\ref{eq:5.2}) obeys the hypotesis (\ref{eq:5.1}) of the theorem of
central limit. In fact a much slower decrease would be enough.

If, for the sake of the argument, we assume that $\pi(y)$ is gaussian, then we
easily compute, by use of Eq.(\ref{eq:5.2})
\begin{eqnarray}
\langle \mu\rangle &=& \bar\mu\exp(\frac{\sigma_y^2}{2})\label{eq:5.3}\\
\sigma_\mu &=& \bar\mu\exp(\sigma^2_y)\nonumber\end{eqnarray}
Eq.'s(\ref{eq:5.3}) show why a direct determination of $\langle\mu\rangle$ is affected
by wild fluctuations: the width is indeed bigger than the value of $\langle
\mu\rangle$ itself.
The exponential dependence on $S'$ strongly distorts the distribution when going
from $S'$ to $\mu$.

The histogram of the values of $\mu$ is related to the constrained potential by the
relation\cite{10,19}
\[\exp\left(- V(\Phi)\right) =
\int[{\cal D}\theta] \exp(\beta S) \delta(\mu - \Phi)\]
$V(\Phi)$ has a minimum at $\langle\beta S'\rangle + \frac{\sigma^2_y}{2}$.

If instead we construct the histogram of $\beta S'$ itself, 
the minimum will appear at $\langle\beta S'\rangle$
which is displaced  by
$\frac{\sigma^2_y}{2}$
with respect to the real minimum.

The problem is that the histogram in $\mu$ is exponentially large to fill adequately,
since $\mu$ fluctuates on an exponential scale (typical 
values of $\mu$ on a configuration
 for a reasonable
lattice size range from $10^{150}$ to 0).
A histogram of $\log\mu$, i.e. of $\beta S'$ is easier to compute.

However to go back to the distribution in $\mu$, i.e. to compute
$\langle
\mu\rangle$ and $\sigma_\mu$, 
we must know the distribution $\pi(y)$ with great precision.
In the gaussian approximation the solution is given by Eq.(\ref{eq:5.3}).
A cluster expansion can be
attempted, to evaluate non gaussian effects, but the problem is only shifted. Higher
cumulants of $\pi(y)$ are more and more noisy to determine numerically, and the
computer time needed becomes comparable to the one needed for the direct
determination of $\langle\mu\rangle$.

Finally a
finite size scaling analysis would be needed, analogous to what we did in sect.4.

This is to justify why we used $\rho$ to extract information on the phase
transition, instead of $\langle\mu\rangle$ itself, or of its effective
potential.

The problem is currently under further study.

\end{document}